\documentstyle{article}

\begin{document}
\begin{titlepage}
\renewcommand{\thefootnote}{\fnsymbol{footnote}}
\renewcommand{\baselinestretch}{1.3}
\hfill  TUW - 92 - 07 \\
\vfill

\begin{center}
{\LARGE {All Symmetries} \\ \medskip
{ of Non-Einsteinian Gravity in $d =2$}}
\vfill

\renewcommand{\baselinestretch}{1}
{\large {Thomas Strobl\footnote{e-mail: tstrobl@email.tuwien.ac.at} \\ \medskip
Institut f\"ur Theoretische Physik \\
Technische Universit\"at Wien\\
Wiedner Hauptstr. 8-10, A-1040 Wien\\
Austria\\} }
\end{center}
\vfill

\renewcommand{\baselinestretch}{1.3}
\begin{abstract}
The covariant form of the field equations for two--dimensional $R^2$--gravity
with torsion as well as its Hamiltonian formulation are shown to suggest the
choice of the light--cone gauge. Further a one--to--one correspondence between
the Hamiltonian gauge symmetries and the diffeomorphisms and local Lorentz
transformations is established, thus proving that there are no hidden local
symmetries responsible for the complete integrability of the model. Finally the
 constraint algebra is shown to have no quantum anomalies so that
Dirac's quantization should be applicable.
\end{abstract}
\vfill

\hfill Vienna, June 1992  \\
\pagebreak
\end{titlepage}
\renewcommand{\baselinestretch}{1}

\section{Introduction}

\renewcommand{\thefootnote}{\alph{footnote}}

One of the  unsolved problems of  quantum field theory is to find
a quantized version of general relativity.
A promising step in this direction is the investigation of some
 simpler models in lower dimensions, which was one of
the reasons$^{1,2}$ for considering the Lagrangian
\begin{equation}
{\cal{L}} = - \, e \, ( \frac{\gamma}{4} \, R^{ab\mu\nu} \, R_{ab\mu\nu} +
\frac{\beta}{4} \, T^{a\mu\nu} \, T_{a\mu\nu} + \lambda)
\label{1}                                              %1
\end{equation}
in $d = 2$. (\ref{1}) is an example of a theory with quadratic
terms in curvature and torsion. In two dimensions  it is the
{\em unique} diffeomorphism invariant action, up to surface
terms, yielding second order differential equations for the
zweibein $e_\mu{}^a = g_\mu{}^a$ and the spin connection
$\omega^a{}_{b\mu}$.  Supplementing it e.g. by the
Einstein--Hilbert term $e R$  does not change the classical
field equations since it is a trivial total divergence here.  In
Katanaev et. al.$^1$  the integrability of (\ref{1}) was proven
using the conformal gauge, thereafter Kummer et. al.$^2$ showed
that the light--cone (LC) gauge enables one to find the explicit
form of the general solution.  Nevertheless, although the LC
gauge is  ideally apt to solve the field eqs.  (cf. also sec.\,2
below), the study of properties concerning geodesics, such as
global completeness$^3$, is preferably done within the conformal
gauge.

It is one of the purposes of this paper  to show that the choice
of the LC gauge, as intuitively introduced in Kummer et.
al.$^2$, is almost compelling when starting from the covariant
form of the field equations (sec.\,2) and even more so when
using the Dirac--Hamiltonian formulation (sec.\,3). As a
by--product of this we will find that there is no direct
analogue of decoupling field eqs.  for the Euclidian version of
the theory, except possibly when using complex coordinates on an
intermediary level.  In contrast to previous work$^{2,4}$ we
shall try to stay as covariant as possible in all of our
calculations, keeping e.g. the covariance in the Lorentz indices
within the Hamiltonian formulation. In sec.\,3 we also propose a
solution to the problem of surface terms on the Hamiltonian
level$^{11}$, at least for the case of this model.

The integrability of the system
inevitabely raises the question of symmetries. In a recent publication$^4$
it was related to the fact that the
Poisson bracket relations between the first class (FC) constraints and
the momenta may be written as a deformed $iso(2,1)$--algebra, which
was said to correspond to some 'novel symmetry' visible only on the
Hamiltonian level.

It is, however, the point of view of the present  paper that the
Poisson bracket relations between the constraints (and momenta)
{\em alone} are of little relevance for the integrability and
the symmetries of the system. Due to a general theorem$^5$ it is
always possible to reformulate the first class constraints such
that their Poisson brackets vanish (strongly).\footnote{Such
an abelianization of the contstraints of the present model has been actually
accomplished in Ref.\ 7.}
Then the
information about the symmetries is  completely hidden in the
specific form of the constraints as functions on  phase space.
To our mind the real content of the symmetries corresponding to
FC constraints becomes visible only when being translated to the
Lagrangian level.

In sec.\,4 we will show that there is a one--to--one
correspondence between the symmetry transformations generated by
the FC constraints on the Hamiltonian level and the
diffeomorphisms and local Lorentz transformations leaving the
action (\ref{1}) invariant. With the results of Henneaux et.
al.$^6$ it follows, moreover, that there are {\em no} further
local symmetry transformations hidden in the model.

The Hamiltonian formulation of sec.\,3 also serves as a preparation for a
non--perturbative quantization of the model.  In sec.\,5 we will show that the
constraint algebra has no quantum anomalies.  Therefore Dirac's procedure of
imposing the constraints on a Hilbert space should be applicable. Some of the
related problems will be addressed in sec.\,6; they will be tackled in a
further publication$^7$.

\section{Covariant Field Equations and Light--Cone   Gauge}

%\doppelnummer

Let us first fix some of the notation: An index from the middle
of the Greek alphabet ($\mu, \, \nu, \ldots$) shall correspond
to a holonomic frame $\partial_\mu$, one of the beginning
($\alpha, \, \beta, \ldots$) to an arbitrary frame $e_\alpha$,
and a Latin index ($a, \, b, \ldots$) to an 'orthonormal' frame
$e_a$, i.e. a frame fulfilling  $g(e_a,e_b) =  g_{ab}$ in which
we will restrict $g_{ab}$ only  to be constant and 'normalized'
to $\det \, g_{ab} = (-1)^M$ ($M = 1$ in the Minkowskian case).
The values of the world--indices $\mu$ are taken from $\{0,1\}$,
whereas those of  Lorentz--indices $a$ from $\{ \hat{0},\hat{1}
\}$; for the special case that $g_{ab}$ is of the form
(\ref{A2}) below we write also $\{ +,- \}$ instead of $\{
\hat{0},\hat{1} \}$.  The  $\varepsilon$--tensor (not
pseudo--tensor) is defined by $\varepsilon_{\alpha\beta} =
(-1)^{O_\alpha} \: \sqrt{\mid \det g_{\alpha\beta}
\mid} \, \varepsilon (\alpha\beta)$, $\varepsilon (\alpha\beta)$ being the
$\varepsilon$--symbol ($\varepsilon (01) = 1$ etc.) and
$(-1)^{O_\alpha}$ the orientation of the frame $e_\alpha$. For
convenience restricting ourselves to a positively oriented
orthonormal frame $e_a$ in the following, we have
$\varepsilon_{ab} =
\varepsilon (ab)$, $\varepsilon^{ab} = (-1)^M \, \varepsilon (ab)$,
$\varepsilon_{\mu\nu} = e \, \varepsilon (\mu\nu)$, and
$\varepsilon^{\mu\nu} = (-1)^M \, (1/e) \, \varepsilon
(\mu\nu)$, the sign of $e$ giving the orientation of
$\partial_\mu$. Furthermore, we will set $\omega_{ab\mu} =:
\varepsilon_{ab} \, \omega_\mu$, which now is invariant only
against global Lorentz--transformations, but the full covariance
can be regained immediately by the substitution $\omega_\mu =
(-)^M \, (1/2) \
\varepsilon^{ab} \, \omega_{ab\mu}$.

The first step in order to find the covariant form of the field
equations is to express (\ref{1}) in terms of the curvature
scalar $R = R^{\alpha\beta}{}_{\alpha\beta}$ and the Hodge dual
of the torsion two--form $T^a =
\frac{1}{2} \,\varepsilon^{\mu\nu} \, T^a{}_{\mu\nu}$. This is possible in
$d = 2$ since one has e.g. $R_{\alpha\beta\gamma\delta} = (-1)^M
\, (R/2) \,
\varepsilon_{\alpha\beta} \, \varepsilon_{\gamma\delta}$ following from the
symmetries of the curvature tensor and  the frequently used
contractions of $\varepsilon^{\alpha\beta} \,
\varepsilon_{\gamma\delta} = (-)^M \, (\delta^\alpha{}_\gamma \,
\delta^\beta{}_\delta - \delta^\alpha{}_\delta \, \delta^\beta{}_\gamma )$.
Thus one obtains
\begin{equation}
{\cal L} \, = \, - \, e \, [\frac{\gamma}{4} \, R^2 + (-1)^M \,
\frac{\beta}{2} \,T^2 + \lambda]
\label{Lneu}                                                %Lneu
\end{equation}
with $T^2 \equiv T^c \, T_c$. Using Cartan's structural
equations, the second one of which is $R = -2 \,
\varepsilon^{\mu\nu} \, \omega_{\mu},_\nu$ in $d = 2$, and
relations such as $\delta e = e \, e_a{}^\mu \, \delta
e_\mu{}^a$, $e_a{}^\mu = g_a{}^\mu$ being the inverse of
$e_\mu{}^a$, the variational principle yields for (\ref{Lneu})
(covariant derivatives are denoted by indices after a
semicolon):
\begin{eqnarray}
 \beta \, T_{a;b}&=& \varepsilon_{ab} \, E(R,T^2)
\label{2a}  \\                                                    %2a
\gamma \, R;_a &=& (-)^M \, \beta \, \varepsilon_{ab} \, T^b
 \label{2b}                                                      %2b
\end{eqnarray}
with
\begin{equation}
E \; \equiv \; \frac{\gamma}{4} \, R^2 +
\frac{(-)^M \beta}{2} \, T^2 - \lambda.
\label{3}                                                        %3
\end{equation}
As a natural extension of the usual variational principle in classical point
mechanics we have required the variation of the fields $e_\mu{}^a$ and
$\omega_\mu$ to vanish at the boundary $\partial B$ of the parameter area $B$
so that  we were allowed to drop the surface term
\begin{equation}
\int_{\partial B} \, e \, \varepsilon_\mu{}^\nu \, [- \gamma
\, R \, \delta \omega_\nu + (-)^{M+1} \beta \, T_a
\delta e_\nu{}^a] \, dx^\mu .
\label{3b}                                                          %3b
\end{equation}
Since (\ref{2a}) and (\ref{2b}) are tensor equations all Latin indices can
be replaced by Greek ones. The field equations have to be
supplemented by the additional geometric requirement $e \equiv \det
e_\mu{}^a \neq 0$.

The equations (\ref{2a}), (\ref{2b}) {\em immediately} show that for
$\mbox{sgn}(\lambda) = \mbox{sgn}(\gamma)$ there exists
a solution with $T^a \equiv 0$ and
$R \equiv \pm \sqrt{4 \lambda / \gamma}$, which has been called
the deSitter solution$^2$.

In order to find the complete solution, we
 write the equations of motion (e.o.m.) explicitely in  the $e_a$--basis,
still for  an arbitrary
(constant and normalized) reference metric $g_{ab}$ :
\begin{eqnarray}
 & T_{\hat{1};\hat{1}} = T_{\hat{0};\hat{0}} = 0 &
\label{4}      \\                                                  %4
& - T_{\hat{0};\hat{1}} = T_{\hat{1};\hat{0}} = - \frac{\textstyle
1}{\textstyle \beta}\, E(R,T^2) &
\label{5}      \\                                                  %5
& R,_{\hat{0}} = \frac{\textstyle  \beta}{\textstyle \gamma}
(- g_{\hat{0}\hat{1}} \, T_{\hat{0}} +
 g_{\hat{0}\hat{0}} \, T_{\hat{1}} ) &
\label{6}      \\                                                  %6
& R,_{\hat{1}} =
\frac{\textstyle  \beta}{\textstyle \gamma}
(- g_{\hat{1}\hat{1}} \, T_{\hat{0}} +
 g_{\hat{0}\hat{1}} \, T_{\hat{1}} ) &
\label{7}                                                          %7
\end{eqnarray}
To integrate at least part of the above equations directly, it is
suggestive to replace a covariant derivative, say '$;_{\hat{0}}$',
by a normal coordinate derivative ('$\partial_0$'). That such a gauge is
attainable  at all --- in $d = 2$ and the covariant derivative
$;_{\hat{0}}$ following Lorentz indices ---
shall be shown in the Appendix A. There we will see also that
we cannot  choose this gauge
in a topology of a torus (staying within one chart by the
requirement of periodic continuation).
Furthermore, for the topology of a cylinder the above
coordinate $x^0$ has to be the one $\in R$. Now, of the  eqs.
(\ref{4}) to (\ref{7})
there are three which are to be integrated for $x^0$. The first one gives
$T_{\hat{0}} = - (\gamma / \beta) A(x^1)$
with an arbitrary  function A. To integrate (\ref{6}) next
(and thereafter the second equation of (\ref{5})), one obviously has to choose
$g_{\hat{0}\hat{0}} = 0$.
Staying with a real tangent space, this can only be achieved for the
case $M=1$. Thus, at least without introducing complex coordinates, the
field eqs. corresponding to the Euclidean version of (\ref{1}) cannot be
decoupled as easily as in the Minkowskian case. Therefore in the
following we will consider only $M = 1$.

What we have demanded so far is already equivalent to the LC gauge as
introduced in Kummer et. al.$^2$.
To show this, we first note that the coordinate
change$^{2,4}$  \, $x^{\pm} = (1/\sqrt{2}) (x^0 \pm x^1)$,
performed  before having introduced any gauge, comes down to a mere relabelling
(since it is a diffeomorphism). Further one notices that the above requirement
$e_{\hat{0}}{}^\mu = \delta (0\mu)$ implies  for
the inverse matrix, i.e. the zweibein,  $e_0{}^a = \delta (\hat{0}a)$
and vice versa. This last equation still
corresponds to a reference metric with an arbitrary $g_{\hat{1}\hat{1}}$, but
there {\em exists} a (global) frame transformation from this $g_{ab}$ to a
light--cone metric
\begin{equation}
g_{ab} =\left( \begin{array}{cc}  0 &  1 \\  1 & 0 \end{array}  \right)
\label{A2}                                              %A2
\end{equation}
so that  $e_0{}^a = \delta (\hat{0}a)$ remains unchanged. This would be e.g.
not the case, if it originally   corresponded to $g_{ab} = {\rm diag}(1,-1)$.
Finally, $\omega_{ab0} =0$ is invariant under any global frame transformation
(cf. (\ref{54}) below). So, the natural requirement of replacing a covariant
derivative by a normal one  and the requirement
of an obvious decoupling of the covariant field eqs. (\ref{2a}),(\ref{2b}) or
(\ref{4}) to (\ref{6})
{\em automatically leads to} the intuitively introduced light--cone gauge of
Kummer et. al.$^2$.

At this point let us observe that the above gauge already
fixes the space--time character of the coordinate $x^0$ to be a light--cone
coordinate: $g_{00} = e_0{}^a \, e_0{}^b \, g_{ab} = g_{\hat{0}\hat{0}} = 0$.
So when referring the zweibein to (\ref{A2}), it is not so severe that the
LC gauge is not attainable for $x^0 \in S^1$, since this corresponds to an
'unphysical world' anyway.
$g_{11} = 2 e_1{}^{\hat{0}}\,e_1{}^{\hat{1}} + (e_1{}^{\hat{1}})^2 \,
g_{\hat{1}\hat{1}}$, on the other hand,
is to be determined by the field eqs. and in general will differ from zero.
%Therefore in a region where $g_{11} < 0$ it might be possible to
%turn Poisson brackets at
%fixed evolution parameter $x^0$ consistently into corresponding commutators
%(cf. sec.\,3 and sec.\,5 below) without getting causality problems.

For reasons of completeness and because our formulation differs
from  previous work, we will briefly sketch how
to solve the remaining field equations.
Since there is no sense in keeping $g_{\hat{1}\hat{1}}$ arbitrary for this
purpose, the zweibein will refer to the  LC  metric (\ref{A2}) for the rest
of this section. Knowing the $x^0$--dependence of $T^{-} = T_{+}$
as well as of $R$ and $T^{+}$, one first determines the dependence of
$e_{-}{}^\mu = (1/e_1{}^{-}) (- e_1{}^{+}, 1)$ and
$\omega_{1}$ on this coordinate, in order to reformulate
the remaining eqs. of (\ref{4}) to (\ref{7}) as ordinary differential eqs.
in $x^1$. To this end one uses the defining eqs. for torsion and curvature
which in this gauge are $T^a = - (1/e_1{}^{-}) [ \partial_0 e_1{}^a +
\delta(a+) \, \omega_1]$ and $R = -2 (\partial_0 \omega_1)
/e_1{}^{-}$.  Thus one obtains for $A(x^1) \neq 0$:
\begin{eqnarray}
R &=& A \, x^0 + B +  \frac{\textstyle \beta}{\textstyle \gamma}
\nonumber \\
T^{+} &=& C \, \exp(-
\frac{\textstyle \gamma}{\textstyle \beta} \, A \, x^0) -
\frac{\textstyle 1}{\textstyle 4A} (A \, x^0 + B)^2 +
\frac{\textstyle \lambda - (\beta^2/4\gamma)}{\textstyle \gamma A}
\nonumber \\
T^{-} &=& - \frac{\textstyle \gamma}{\textstyle \beta} \, A
\nonumber \\
 \omega_1 &=& F - \frac{\textstyle \beta}{\textstyle 2\gamma} \,
\frac{\textstyle D}{\textstyle A} \,
\exp(\frac{\textstyle \gamma}{\textstyle \beta} \, A \, x^0) \, (A\, x^0 + B)
\nonumber \\
e_1{}^{+} &=& G - (F + CD) \, x^0 +
\frac{\textstyle \beta \, D}{\textstyle 4\gamma \, A^2}
\, [(A \, x^0 + B)^2 + \frac{\textstyle \beta^2}{\textstyle \gamma^2} -
\frac{\textstyle 4 \lambda}{\textstyle \gamma}
] \, \exp(\frac{\textstyle \gamma}{\textstyle \beta} \, A \, x^0)
\nonumber \\
e_1{}^{-} &=& D \, \exp(\frac{\textstyle \gamma}{\textstyle \beta} \, A \, x^0)
\label{8}                                                      %8
\end{eqnarray}
with $A, B, C, D, F, G$ being functions of $x^1$ ($D(x^1) \neq 0$ because
of $e_1{}^{-} = e$). Likewise for $A(x^1) = 0$ one gets:
 \begin{eqnarray}
R &=& B +  \frac{\textstyle \beta}{\textstyle \gamma}
\nonumber \\
T^{+} &=& \tilde{C} + I \, x^0
\nonumber \\
T^{-} &=& 0
\nonumber \\
 \omega_1 &=& \tilde{F} - \frac{\textstyle \beta}{\textstyle 2\gamma} \,
D \, (1+ \frac{\textstyle \gamma}{\textstyle \beta} \, B ) \,  x^0
\nonumber \\
e_1{}^{+} &=& \tilde{G} - (\tilde{F} + \tilde{C}D) \, x^0 +
\frac{\textstyle D}{\textstyle 4} (B +
\frac{\textstyle \beta}{\textstyle \gamma} - I) \, (x^0)^2
\nonumber \\
e_1{}^{-} &=& D
\label{12}                                                        %12
\end{eqnarray}
with
\[
I \;\equiv \;
\frac{\lambda}{\beta}
- \frac{\gamma}{4\beta} (B + \frac{\beta}{\gamma})^2
\]
and with the  arbitrary functions $B, \tilde{C}, D \neq 0, \tilde{F},
\tilde{G}$.
%The comparison of (\ref{8}) with (\ref{12})
%reveals that if $A(x^1) \neq 0$ at some point $x^1$ it has to stay different
%from zero everywhere. The possibility  $\lambda = (\beta^2/
%4\gamma)$ (and $B \rightarrow 0$  when $A \rightarrow 0$ etc.) is ruled
%out also by the remaining eqs. of (\ref{4}) to (\ref{7}), at least when
%assuming that the metric and the connection
%are continous in $x^1$, thus excluding coordinate singularities.
% Note that
%e.g. equation (\ref{16}) below shows that with $A \rightarrow 0$ also
%$A^{(n)}
%\rightarrow 0$.
For the case $A = 0$ everywhere the remaining eqs.  of (\ref{4}) to (\ref{7})
can be shown to lead
to $B = \pm \sqrt{(4\lambda/\gamma)} - (\beta/\gamma)$ ($\Leftrightarrow I =
0$) and $\tilde{C} = 0$ everywhere, which is the deSitter part of the
solution.
For $T^{-} \neq 0$ everywhere it suffices to
regard only the first eq. of (\ref{4}), the other two eqs. being redundant.
It yields in this case (the prime denoting a derivative):
\begin{eqnarray}
&A' + A (CD + F) = 0&
\label{16}      \\                                                  %16
&B' - \frac{\textstyle \beta}{\textstyle \gamma} \, CD - AG = 0&
\label{17}      \\                                                  %17
&C' - C (F - \frac{\textstyle \gamma}{\textstyle \beta} \, AG) = 0&
\label{18}                                                          %18
\end{eqnarray}
Plugging (\ref{16}, \ref{17}) into (\ref{18}) one can easily integrate
the latter equation to give
\begin{equation}
C = \frac{C_1}{A} \, \exp [-\frac{\gamma}{\beta} \, B  - 1]
\label{19}                                                          %19
\end{equation}
for an arbitrary constant $C_1$. Eqs. (\ref{16}) and (\ref{17}) finally can be
solved by expressing $F$ and $G$ in terms of the remainig three functions.

Calculating the Lorentz invariant $T^2 = 2 \ T^{+}
\, T^{-}$ and making the result explicit in $C_1$, one regains the
quantity $Q(R,T^2) \propto  C_1$
\begin{eqnarray}
Q &=& \exp (\frac{\textstyle \gamma}{\textstyle \beta} \, R) \,
[\frac{\textstyle 2\gamma}{\textstyle \beta} \, T^2 -
\, (\frac{\textstyle \gamma}{\textstyle \beta} R- 1)^2 -1 +
\frac{\textstyle 4\gamma\lambda}{\textstyle \beta^2}
 \; \equiv \nonumber \\
& \equiv & 2 \exp (\frac{\gamma}{\textstyle \beta} \, R) \,
[- \frac{\textstyle 2\gamma\lambda}{\textstyle \beta^2} E +
\frac{\textstyle \gamma}{\textstyle \beta} \, R  -1]
\label{19b}                                                          %19b
\end{eqnarray}
of Kummer et. al.$^2$ (up to a proportionality constant).
Being a covariant function, we have gauge--independently
\begin{equation}
Q;_a \, = \, 0,
\label{19c}                                                          %19c
\end{equation}
 which also follows immediatly from taking the covariant derivative
 of the first equation of (\ref{19b}) and the subsequent usage of (\ref{2a}),
  (\ref{2b}).

There exist also solutions with isolated zeros of $A(x^1)$ as
 outlined in Ref.\,7.

\section{Hamiltonian Formulation}

%\doppelnummer

Again $g_{ab}$ will be
restricted only by $g_{ab} = const$ and $\det g_{ab} = -1$ in the following.
Further we will use  the notation  $\partial_0 f =: \dot{f}, \: x^1 =: x,
\: \partial_1 f =: \partial f$ and  choose,  without loss of generality,
$x^0 =: t$  as evolution parameter; clearly
 the classical  system is reproduced
by the Hamiltonian formalism irrespective of the space--time character of $t
\equiv x^0$, which is {\em not} fixed at this stage.
We observe first that  (\ref{1}) or (\ref{Lneu}) is of the simple form
\begin{equation}
{\cal L} = \frac{1}{2} \, (\dot{\varphi}^i - f^i) \, W_{ij} \,
(\dot{\varphi}^j - f^j) - V
\label{20}                                                            %20
\end{equation}
with
\begin{eqnarray}
& \varphi^A \equiv  (\varphi^i ,
\bar{\varphi}^i) \equiv (e_1{}^a,\omega_1,
e_0{}^a,\omega_0) \qquad i = (\hat{0},\hat{1},\omega), &
\nonumber            \\
\vspace{6ex}
&f^i \equiv
(\dot{e}_1{}^a - T^a{}_{01}, \partial \omega_0) =
f^i(\varphi^A,\partial \bar{\varphi}^j) &
\nonumber             \\
\vspace{3ex}
& V \equiv \lambda \, e, \quad
W_{ij}  \equiv \frac{\textstyle 1}{\textstyle e} \left(
\begin{array}{cc} \beta \,g_{ab} & 0 \\ 0 &  -2 \gamma \end{array}
\right).&
\label{23}                                                          %23
\end{eqnarray}
Therefore ${\cal L}$ is completely regular in the three fields $\varphi^i$
--- i.e. the Legendre transformation is bijective in these fields ---
and completely singular in $\bar{\varphi}^i$. So the Legendre
transformation in the regular sector $\pi_i \equiv (\pi_a, \pi_\omega) = W_{ij}
(\dot{\varphi}^j - f^j) = (-\beta \, T_a, \gamma \, R)$  gives just the inverse
to the Hamiltonian equations for   $\dot{\varphi}^i$ (cf. eq. (\ref{35})
below) and  the primary constraints are simply $\bar{\pi_i} \approx 0$. Since
we have $(W^{-\!1})^{ij} = e \,
\mbox{'diag'}(g^{ab}/\beta, -1/2\gamma)$, the canonical Hamiltonian
\begin{equation}
{\cal H}_C = \frac{1}{2} \, \pi_i \,
(W^{-\!1})^{ij} \, \pi_j + f^i \, \pi_i + V
\label{24}                                                          %24
\end{equation}
is polynomial in the basic fields, in contrast to the Lagrangian and
also in contrast to the conventional Hamiltonian  of general relativity
in $d = 4$ (but similar to the Ashtekar formulation of the latter ---
cf. Isham$^{12}$).
This seems to indicate an  advantage of the Hamiltonian path integral
formulation as opposed to the Lagrangian one.
However, when formally integrating out the momenta$^8$
of a BRST--version
of the canonical Hamiltonian (with $\bar{\varphi}^i$ interpreted as Lagrange
multipliers -- cf. comments following (\ref{33}) below),
one just regains\footnote{Up to a factor in the measure of the
Lagrangian path integral which basically originates from the fact
that the Hessian  (3.2) is not constant.}
the Lagrangian as used in Kummer et. al.$^9$;
the four--vertex ghost couplings$^{10}$, typical for theories
with a non--closing constraint algebra,  disappear in this model.

To get e.o.m. and constraints which  are equivalent to
the Lagrangian field eqs., one has to use  the primary Hamiltonian
${\cal{H}}_P$, which is the sum of
${\cal{H}}_C$ and the product of all primary
constraints with  Lagrange multiplier (LM) functions. Usually this is proven
for discrete systems (cf. Sundermeyer$^{11}$ and references therein),
and a formal translation to the continuous
case would imply  ${\cal{H}}_P = {\cal{H}}_C + \bar{\lambda}_i \,
\bar{\pi}_i$. But in our opinion this is not enough to reproduce exactly what
we have done on the Lagrangian sector (sec.\,2).
There we have required the variation of the
basic fields to vanish at the boundary of the parameter
space B in order to get
the field equations (\ref{2a}), (\ref{2b}).
This can be viewed as a minimalization of the
corresponding action with 'temporarily fixed'
boundary conditions on $\partial$B, finally asking for the set of all
solutions compatible with {\em some} boundary condition.
Now, any of the usual proofs working in a discrete version of a continuous
Lagrangian system correspond to a {\em rectangular} B $\subset R^2$.
But for such a B  the
'temporary' fixation of the
fields at $t_{min}$ and $t_{max}$ have to be supplemented by 'temporary'
boundary conditions for the field variables at $x_{min}$ and $x_{max}$.
According to (\ref{3b})  it suffices to
prescribe such boundary conditions only
for the $\varphi^i$ (and not necessarily for the
$\bar{\varphi}^i$) in our case;
they are explicitly $t$--dependent and have to be added as
{\em additional primary constraints} to the Hamiltonian formulation of the
system.

As a next step one has to calculate the secondary constraints.  Considering
$\varphi^i \! \mid_{x-boundary} \,  \approx \, g^i(t)$ first, the requirement
that its Poisson brackets with $H_P \equiv \int dx \, {\cal
H}_P$ shall vanish reproduces just $\pi_i \! \mid_{x-boundary}
\: \approx (-\beta \, T_a, \gamma
\, R)\! \mid_{x-boundary}$ (cf. (\ref{24})) when identifying $\dot{\varphi}^i
\! \mid_{x-boundary}$ with $\dot{g}^i(t)$. All further tertiary, etc.
constraints can be reformulated to be only restrictions to the
possible choice for $g^i(t)$.  So we are left with a set of
$t$--dependent {\em second class} constraints at the boundary.
The corresponding degrees of freedom can be eliminated by means
of Dirac brackets so that we will be allowed to set all surface
terms strongly equal to zero below.

A  simple calculation,
using  $e = \varepsilon(ab) \, e_0{}^a \, e_1{}^b$,
 shows that the conservation of $\bar{\pi}_i \approx
0$ gives ($G_i := \{ \bar{\pi}_i, H_P \}$):
\begin{eqnarray}
G_a &=& \varepsilon_{ab} \, e_1{}^b \, E +
\partial \pi_a - \varepsilon^b{}_a \,
\omega_1 \, \pi_b \approx 0
\label{25}            \\                                              %25
G_{\omega} &=& \partial \, \pi_{\omega} + \varepsilon^a{}_b \, e_1{}^b \, \pi_a
\approx 0
\label{26}                                                         %26
\end{eqnarray}
with
\begin{equation}
 E \; \equiv \; \frac{1}{4\gamma} \, (\pi_{\omega})^2
  - \frac{1}{2\beta} \, \pi^2 -
\lambda, \quad \pi^2 \, \equiv \, \pi^a \, \pi_a.
\label{27}                                                            %27
\end{equation}
\begin{sloppypar}
On shell (using the e.o.m. for $\dot{\varphi}^i$)
(\ref{27}) becomes just (\ref{3}). It should be
mentioned, furthermore, that on
deriving (\ref{25}, \ref{26}) we have replaced
$\int dy \, \partial_y [\delta(y - x)] \, \pi_i(y)$ by
$-\partial \pi_i(x)$, which is possible only since $\pi_i$ vanishes strongly
at $x_{min}, x_{max}$.
\end{sloppypar}

At this point it is convenient  to calculate the
brackets of the basic fields with the $G_i$; the only nonvanishing ones  are
(we suppress the arguments $x$ and $y$ and  write
$\delta '$ for $\partial \delta (x - y)$ etc.):
\begin{eqnarray}
&\{ e_1{}^a, G_b \} =
(\frac{\textstyle -1}{\textstyle \beta} \,
\varepsilon_{bc} \, e_1{}^c \,\pi^a
-  \varepsilon^a{}_b \, \omega_1) \, \delta - \delta^a{}_b \, \delta ', \quad
\{ e_1{}^a, G_{\omega} \} = \varepsilon^a{}_b \, e_1{}^b \, \delta \;&
\label{28}            \\                                              %28
&\{ \omega_1, G_a \} = \frac{\textstyle 1}{\textstyle 2\gamma} \,
\varepsilon_{ab} \, e_1{}^b \,
\pi_{\omega} \, \delta , \quad \{ \omega_1, G_{\omega} \} = - \delta '&
\label{29}            \\                                              %29
&\{ \pi_a, G_b \} = \varepsilon_{ab} \, E \, \delta, \quad
\{ \pi_a, G_{\omega} \} = \varepsilon_{ab} \, \pi^b \, \delta, \quad
\{ \pi_{\omega}, G_a \} = - \varepsilon_{ab} \, \pi^b \, \delta.&
\label{30}                                                            %30
\end{eqnarray}
By means of this one derives
\begin{eqnarray}
\{ G_a, G_{\omega} \} &=& -  \varepsilon^b{}_a \, G_b \, \delta
\label{31}            \\                                              %31
\{ G_a, G_b \} &=& \varepsilon_{ab} \,(
\frac{\textstyle -1}{\textstyle \beta} \,\pi^c \, G_c +
\frac{\textstyle 1}{\textstyle 2\gamma} \, \pi_{\omega}
\, G_{\omega} ) \, \delta .
\label{32}                                                            %32
\end{eqnarray}
(\ref{30}) to (\ref{32})  is the Lorentz--covariant version of the algebra
analyzed in Grosse et. al.$^{4,\,}$\footnote{
For a detailed comparison one has to set $\gamma = 1/2$, to
replace $\beta$ by $2 \beta$, and to replace $i = (\hat{0}, \hat{1}, \omega)$
or $(+,-,\omega)$ by
$i = (3,2,-1)$ [i.e.\ $\varphi^i$ by $(q_3,q_2,- q_1)$, etc.].}.

Dropping $\partial (\bar{\varphi}^i \, \pi_i)$ since it does not contribute
to $H_P$ by the above argument,  we can express our Hamiltonian by means of
(\ref{24}) to (\ref{27}) as
\begin{equation}
{\cal H}_P = - \bar{\varphi}^i \, G_i(\varphi,\pi,\partial \pi) +
\bar{\lambda}^i \, \bar{\pi}_i.
\label{33}                                                            %33
\end{equation}
If we had not employed the argument with the Dirac brackets, (\ref{33}) would
be still valid since, as alluded to above, one would have to include the
(not completely well--defined) term $- \delta(x - x_{boundary}) \,
\pi_i(x)$ in the formulas (\ref{25},\ref{26}) for the secondary constraints
$G_i$.

%\begin{sloppypar}
That the Hamiltonian vanishes weakly is a feature in common to all
parame\-trization invariant theories$^{11}$. In this specific case, though,
the structure of (\ref{33}) reveals that from the {\em Hamiltonian} point of
view one could also consider ${\cal H} = \mu_i \, G_i$ instead of ${\cal H}_P$,
which is obtained from (\ref{33}) in the gauge
$\bar{\varphi}^i \approx -\mu^i$,
dropping the barred fields by the introduction
of Dirac brackets.
This is also completely analoguos to the usual Hamiltonian formulation of
general relativity in four dimensions (cf. e.g. Isham$^{12}$), where one
regards
the lapse function and the shift vector as mere Lagrange multipliers.  Because
of (\ref{33}) the relations (\ref{31},\ref{32}) show that there are no further
secondary constraints. Moreover, all of the $6 \, \infty$ constraints
$(G_i,\bar{\pi}_i)
=: G_A$ are first class (FC).
%\end{sloppypar}

Before turning to the symmetries generated by the FC constraints,
let us comment on the flow generated by (\ref{33}), i.e. the
'$t$--evolution' of the coordinates. Supplementing them by the constraints,
we have (cf. (\ref{25},\ref{26})):
\begin{eqnarray}
&\dot{\bar{\varphi}}^i \approx \bar{\lambda}^i, \quad \bar{\pi}_i \approx 0 &
\label{34}            \\                                              %34
&\dot{\varphi}^i \approx \{ \varphi^i, H_P \} \; \Leftrightarrow \;
\pi_i \approx (-\beta \, T_a, \gamma \, R)&
\label{35}                  \\                                          %35
&\partial_{\mu} \, \pi_{\omega} \approx \varepsilon_{ab} \, e_\mu{}^a \, \pi^b&
\label{36}                      \\                                      %36
&\partial_{\mu} \, \pi_a \approx - \varepsilon_{ab} \, ( e_\mu{}^b \, E +
\omega_{\mu} \, \pi^b)&
\label{37}                                                              %37
\end{eqnarray}
Multiplying the latter two eqs. by $e_\mu{}^b$, one obviously regains just
the covariant  e.o.m. (\ref{2a}), (\ref{2b}).

Beside this reformulation of the covariant field equations on the Lagrangian
level the Hamiltonian
formulation (using ${\cal H}_P$) provides some additional information and
insight. On the one hand, the equations (\ref{34}) practically force one to
use a LC--like gauge. Furthermore, the constraint eqs. (\ref{25},\ref{26})
serve as constants of the motion, thus saving one the integration of three of
the e.o.m. (cf. Ref.\,4 or Ref.\,7 for more details).
 On the other hand, the knowledge that (\ref{36}),
(\ref{37}) with $\mu = 1$ are {\em first} class constraints reveals$^{11}$
that {\em on shell} the choice of the LC gauge does not fix the gauge freedom
completely. From the Hamiltonian point of view  the LC gauge turns only
$\bar{\pi}_i \approx 0$ into second class constraints.
%  the barred coordinates
%are then eliminated by the introduction of Dirac brackets.
This  shows
that it should be possible to gauge away still $3 \, \infty$ phase space
coordinates, i.e the three arbitrary functions
of $x \equiv x^1$ in the
general solution obtained at the end of sec.\,2 -- except possibly for a finite
number of constants.  Since we will show in the following section that the
gauge symmetries generated by the FC constraints are just diffeomorphisms and
Lorentz transformations, this elimination could already have
been carried through in sec.\,2. The corresponding steps  on the
Lagrangian level can be found in Grosse et. al.$^4$. According
to their result  $-4(\gamma/\beta)^2 \, C_1 = Q(R,T^2)$ is the only gauge
independent (physical) parameter left in the model. For the
topology of a cylinder, however, the reduced phase space is two dimesional
(cf. Ref.\,7).

One could also try to build up a Hamiltonian formulation with the extended
Hamiltonian ${\cal H}_E$ as proposed by Dirac$^{13}$,
which emerges from ${\cal H}_c$ by adding {\em all}
FC constraints via Lagrange multipliers (LM). When absorbing $-
\bar{\varphi}^i$ into the definition of new LMs $\lambda^i$,
which will prove to be a mathematical shortcut especially in the
following section since it {\em formally} corresponds to a strongly vanishing
canonical Hamiltonian,
one obtains ($\lambda^A := (\lambda^i, \bar{\lambda}^i)$):
\begin{equation}
{\cal H}_E = \lambda^A \, G_A.
\label{38}                                                          %38
\end{equation}
In this way the e.o.m. for the unbarred fields become proportional to LMs,
too.  Setting all the $\lambda^A$ zero as the simplest choice, one is left
with $\varphi^A \approx
\varphi^A (x), \, \pi_A \approx (\pi_i (x), 0)$,  in which the nine functions
of $x$ are restricted by the three contraints $G_i$. The FC character of the
six constraints $G_A$ will  lead to a gauge identification of the six functions
needed as 'initial data' on a hyperline $t = const$ so that  the (superficial)
degrees--of--freedom count gives the same as in the case of ${\cal H}_P$.  That
the fixation of all LMs in ${\cal H}$ does not fix the gauge completely can be
understood by considering a hyperline for some smaller $t$: Starting from such
a line with different choices for the LMs  leads to different but physically
equivalent points on the hyperline used above. Nevertheless, without having
gone into further calculational detail, we  believe that it would be  difficult
to extract geometrically interpretable results comparable to the results of
sec.\,2 when using ${\cal H}_E = 0$ as above. (Note also that the
gauge choice $\lambda^A =0$ in (3.17) corresponds to a gauge with $e=0$ in the
original formulation).
However, for the gauge independent quantity $Q (R,T^2)$ the e.o.m. are
unchanged: $\dot{Q} \approx 0$ is trivial and $\partial Q \approx 0$ follows
from $G_i \approx 0$.

\section{All local symmetries of the model}

%\doppelnummer

Thanks to the work of Henneaux et. al.$^6$ the relationship
between the gauge symmetries on the Hamiltonian level and the
ones on the Lagrangian level has become much more precise than
it had been before (cf. e.g.  Sundermeyer$^{11}$). The
assumptions are quite  general, but excluding e.g. ineffective
constraints so as to evade counterexamples of 'Dirac's
conjecture'$^{13}$ as the one given in Gotay$^{14}$. The main
idea of Henneaux et. al.$^6$ is to restrict the symmetries of
the action $S_E = \int d^2x \, (\dot{\varphi}^A \, \pi_A - {\cal
H}_E)$ successively to the ones of $S_P = \int d^2x \,
(\dot{\varphi}^A \, \pi_A - {\cal H}_P)$ and $S_L = \int d^2x \,
{\cal L}$.   The present section is a nice illustration for the
general considerations made in that work as well as for the
usefulness of ${\cal H}_E$, i.e. of 'Dirac's conjecture', when
analyzing the symmetries of a model.

With
\begin{equation}
\{ G_B(x),G_C(y) \} = \delta(x-y) \, C_{BC}{}^A \, G_A
\label{41}                                                            %41
\end{equation}
it is easy to verify that the gauge transformations
\begin{equation}
 \delta f(x,t) =  \int dy \, \{ f(x,t), G_A(y,t) \} \, \epsilon^A(y,t),
 \label{39}                                                          %39
\end{equation}
in which $f = f(\varphi^A(x,t), \pi_A(x,t))$ is an arbitrary function on
the phase space and $\epsilon^A(x,t)$ an arbitrary infinitesimal parameter,
are a symmetry transformation of $S_E$ (up to surface terms), iff
\begin{equation}
\delta \lambda^A = \dot{\epsilon}^A + C_{BC}{}^A \, \lambda^C \, \epsilon^B.
\label{40}                                                            %40
\end{equation}
If we had not absorbed ${\cal H}_c$ by the definition of the LMs, we had to
add $- V_B^A \,  \epsilon^B$ to (\ref{40}) ($\{ H_c (t), G_A(x,t) \} =: V_B^A
\, G_B$). Now, according to Henneaux et. al.$^6$ these are {\em all} local
transfomations leaving $S_E$ invariant.

In our case  the above equations become (cf. (\ref{28}) to (\ref{30})):
\begin{eqnarray}
\delta \bar{\varphi}^i &=& \bar{\epsilon}^i, \; \quad
\delta \bar{\pi}_i = 0, \;\quad \delta \bar{\lambda}^i =
\dot{\bar{\epsilon}^{\,i}}
\label{42}            \\                                              %42
\delta e_1{}^a &=& - \frac{\textstyle 1}{\textstyle \beta} \,
\varepsilon_{bc} \, e_1{}^c \, \pi^a \,
\epsilon^b - \varepsilon^a{}_b \, \omega_1 \,  \epsilon^b - \partial
\epsilon^a + \varepsilon^a{}_c \, e_1{}^c \, \epsilon^{\omega}
\label{43}         \\                                                 %43
\delta \omega_1 &=& \frac{\textstyle 1}{\textstyle 2 \gamma} \,
\varepsilon_{bc} \, e_1{}^c \,
\pi_{\omega} \, \epsilon^b - \partial  \epsilon^{\omega}
\label{44}         \\                                                 %44
\delta \pi_a &=&  \varepsilon_{ab} \,E\,  \epsilon^b + \varepsilon_{ac}
\pi^c \, \epsilon^{\omega}
\label{45}         \\                                                 %45
\delta \pi_{\omega} &=& - \varepsilon_{bc} \, \pi^c \, \epsilon^b
\label{46}     \\                                                     %46
\delta \lambda^a &=& \dot{\epsilon^a} +
\frac{\textstyle 1}{\textstyle \beta} \, \varepsilon_{bc} \,
 e_0{}^c \, \pi^a \, \epsilon^b + \varepsilon^a{}_b \, \omega_1 \,
 \epsilon^b - \varepsilon^a{}_c \, e_0{}^c \, \epsilon^{\omega}
\label{47}         \\                                                 %47
\delta \lambda^{\omega} &=& \dot{\epsilon^{\omega}} +
\frac{\textstyle 1}{\textstyle 2 \gamma} \,
\varepsilon_{bc} \, e_0{}^c \, \pi_{\omega} \, \epsilon^b
\label{48}                                                            %48
\end{eqnarray}
Setting $\lambda^i = - \bar{\varphi}^i$, the action $S_E$ becomes equal to
$S_P$. Thus, all local symmetries leaving $S_P$ invariant are obtained when
restricting (\ref{42}) to (\ref{48}) by $\delta \lambda^i = - \delta
\bar{\varphi}^i \equiv - \bar{\epsilon}^i$. This gives among others:
\begin{eqnarray}
\delta e_\mu{}^a &=& - \partial_\mu \epsilon^a  -
\frac{\textstyle 1}{\textstyle \beta} \,
\varepsilon_{bc} \, e_\mu{}^c \, \pi^a \,
\epsilon^b - \varepsilon^a{}_b \, \omega_\mu \,  \epsilon^b  +
\varepsilon^a{}_c \, e_\mu{}^c \, \epsilon^{\omega}
\label{49}         \\                                                 %49
\delta \omega_\mu &=& \frac{\textstyle 1}{\textstyle 2 \gamma} \,
\varepsilon_{bc} \, e_\mu{}^c \,
\pi_{\omega} \, \epsilon^b -  \partial_\mu \epsilon^{\omega},
\label{50}                                                            %50
\end{eqnarray}
whereas the transformations for the momenta remain unchanged. The
transition from
$S_P$ to $S_L$ is accomplished by the second equations of (\ref{34}) and
(\ref{35}). Using the latter to eliminate the momenta from (\ref{49}),
(\ref{50}) as well as the identities\footnote{
The brackets $[ \; ]$ indicate antisymmetrization, the comma a partial
derivative: $\omega_{[\nu},_{\mu ]} \equiv (1/2) (\partial_\mu \, \omega_\nu -
\partial_\nu \, \omega_\mu$).}
$(T^a,R/2) \, \varepsilon_{\nu\mu} = (T^a{}_{\mu\nu},2\omega_{[\nu},_{\mu ]})$
 and the abbreviation $e_b{}^\nu \epsilon^b \equiv  \epsilon^\nu$,
one verifies easily:
\begin{eqnarray}
\delta e_\mu{}^a &=& - \partial_\mu \epsilon^a  + 2 e_{[ \nu}{}^a,_{\mu ]}
\epsilon^\nu +
\varepsilon^a{}_c \, e_\mu{}^c \,(\epsilon^{\omega} -  \omega_\nu
\,  \epsilon^\nu)
\label{51}         \\                                                 %51
\delta \omega_\mu &=&  \omega_\mu,_\nu \,  \epsilon^\nu - \omega_\nu
\,  \epsilon^\nu,_\mu  -  \partial_\mu (\epsilon^{\omega} -  \omega_\nu
\,  \epsilon^\nu)
\label{52}                                                            %52
\end{eqnarray}
These are the transformations fulfilling $\delta S_L = 0$.

Before turning to some consistency checks regarding the momenta, let us
verify that (\ref{51}, \ref{52}) are nothing but the infinitesimal form
of diffeomeorphisms and Lorentz transformations. The wellknown parametrization
of the Lorentz group in $1+1$ dimensions with cosh$\alpha$, sinh$\alpha$
$(\alpha = \alpha (x^\mu) \in R)$ gives in its infinitesimal form
\begin{equation}
\delta^L_\alpha \; e_\mu{}^a  = \varepsilon^a{}_b \, \alpha \, e_\mu{}^b.
\label{53}                                                            %53
\end{equation}
The inhomogeneity in $\omega_{\bar{a} \bar{b} \mu} = e_{\bar{a}}{}^c \,
 e_{\bar{b}}{}^d \, \omega_{ab\mu} + e_{\bar{a}c} \, e_{\bar{b}}{}^c,_\mu$,
in which  $e_{\bar{a}}{}^b$ denotes the Lorentz boost inverse to the one used
in (\ref{53}), leads to the transformation property
\begin{equation}
\bar{\omega}_\mu = \omega_\mu - \alpha,_\mu,
\label{54}                                                            %54
\end{equation}
or $\delta^L_\alpha \; \omega_\mu = - \alpha,_\mu$. The usual transformation
under diffeomorphisms $\tilde{x}^\mu := x^\mu + \xi^\mu(x^\sigma)$
for a field $v_\mu$ reads infinitesimally: $\delta^D_\xi \, v_\mu = - v_\nu \,
\xi^\nu,_\mu -  v_\mu,_\nu \, \xi^\nu$.  Since obviously $\delta^D_\xi
\,e_\mu{}^a =  - (e_\nu{}^a \, \xi^\nu),_\mu  - 2 e_{[ \mu}{}^a,_{\nu ]}
\xi^\nu$ the comparison of (\ref{51}, \ref{52}) to the above equations yields:
\begin{equation}
\delta \varphi^A  = (\delta^D_{\epsilon^\nu} + \delta^L_{\epsilon^\omega -
\omega_\nu  \epsilon^\nu}) \, \varphi^A.
\label{57}                                                            %57
\end{equation}
Thus it is found that the flow generated by $G_a$ on the regular sector
is a specific combination of
a diffeomorphism and a Lorentz transformation, whereas
$G_\omega$ corresponds directly to the generator of the
Lorentz group in this sector.

There are no momenta living on the purely Lagrangian sector. Nevertheless,
we do not see a {\em general} reason {\em within} the above procedure\footnote{
Certainly we know already that $S_L$ is invariant under all of (\ref{57}).}
why the  transformation properties of the momenta on the $S_P$--level could
not lead to a restriction of the parameters in (\ref{57}) (similar to the
transition from $S_E$
to $S_P$, in which the number of free parameters has been
halved). To show explicitely that this
is not the case, we note that $\delta \bar{\pi}_i = 0$ is consistent with the
second equation of (\ref{34}). Furthermore, by means of (\ref{35})
the eqs. (\ref{45}) and (\ref{46}) can be written as:
\begin{eqnarray}
\delta (-\beta T_a) &=& [\varepsilon_{ab} \, E(R,T^2) - \beta \,
T_{a;b}] \, \epsilon^b - \beta [ \varepsilon_{ac} \, T^c (\epsilon^\omega -
\omega_\nu \, \epsilon^\nu) - T_a,_\nu \, \epsilon^\nu ] \equiv
\nonumber \\
\nopagebreak
& \equiv &  [\hat{\delta}_{\epsilon^\nu} + \delta^D_{\epsilon^\nu} +
\delta^L_{\epsilon^\omega - \omega_\nu  \epsilon^\nu}]  \: (-\beta T_a)
\label{58}              \\                                            %58
\delta (\gamma R) &=& (\beta \, \varepsilon_{bc} \, T^c
+ \gamma \, R;_b)
\, \epsilon^b - \gamma \, R,_\nu \, \epsilon^\nu \equiv
\nonumber \\
& \equiv &  [\hat{\delta}_{\epsilon^\nu} + \delta^D_{\epsilon^\nu} +
\delta^L_{\epsilon^\omega - \omega_\nu \epsilon^\nu}] \: (\gamma R)
\label{59}                                                            %59
\end{eqnarray}
with $\hat{\delta} = 0$ on shell (cf. (\ref{2a},\ref{2b}))! This completes
the proof for the one--to--one correspondence of the gauge symmetries on
the Lagrangian and the Hamiltonian level, and, as consequence, the absence
of any further local symmetry.

\section{Constraint Algebra without Anomalies}

%\doppelnummer

The first step towards a quantized version of the model is to turn the
$\varphi^A$ and $\pi_A$ into operators on
some --- at this stage unspecified ---
Hilbert space and to replace the fundamental Poisson brackets
$\{ \varphi^A (x), \pi_B (y) \} = \delta^A{}_B \, \delta (x-y)$ by
the commutator relations
$[ \varphi^A (x), \pi_B (y) ] = i \hbar \, \delta^A{}_B \, \delta (x-y)$.
Next, although not mandatory (cf. e.g. Isham$^{12}$), it is  'natural' to
require the constraints $G_A$ to become hermitean operators. This is
accomplished replacing $e_1{}^b  E$ in (\ref{25}) by the anticommutator
$(1/2) \, [e_1{}^b , E]_+$. $G_\omega$ is already hermitean
since we have
\begin{equation}
\varepsilon^a{}_b \,[ e_1{}^b(x) , \pi_a(x)] = \varepsilon^a{}_a \,
i \hbar \,  \delta (0) = 0
\label{60}                                                      %60
\end{equation}
in any  regularization for $ \delta (0)$. To employ Dirac's approach
\begin{equation}
G_A \, \psi_{\mbox{phys}} = 0,
\label{61}                                                      %61
\end{equation}
it is common to demand that one cannot produce further
constraints by applying $G_B$  to the lefthand side of (\ref{61}). This
 is guaranteed, if the constraint
algebra (\ref{31},\ref{32}) has no quantum anomalies, i.e. if
 the quantum version of (\ref{31},\ref{32}) is obtained from these equations
by the mere replacement
\begin{equation}
\{ \, , \, \} \: \rightarrow \: -i/\hbar \, [ \, , \, ]
\label{62}                                                      %62
\end{equation}
without reordering of the (hermitean) operators $G_i$
on the r.h.s. so that some $G_i$
would be placed on the left of $\pi_i$.  To show that there are indeed no
such anomalies for the case of our model, one  first observes
that the constraints are only linear in the fields $\varphi^i$ (cf.
(\ref{25},\ref{26})). It is  easy to verify,
by dropping a similar term
as the one in (\ref{60}), that the commutator between  two operator
valued functions
\[
f = \frac{1}{2} \, [\varphi^i(x) , f_i (\pi_j(x))]_+ \quad {\rm and} \quad
g = \frac{1}{2} \, [\varphi^i(x) , g_i (\pi_j(x))]_+
\]
gives
\begin{equation}
 [f(x),g(y)] = i\hbar \, \frac{1}{2} \, [\varphi^k ,
 (\frac{\partial g_k}{\partial \pi_i} f_i -
 \frac{\partial f_k}{\partial \pi_i} g_i) ]_+ \; \delta (x-y),
\label{63}                                                      %63
\end{equation}
\begin{sloppypar}
which, except for the factor $i\hbar$, is just the antisymmetrization of the
classical result. Applying this  to (\ref{31},\ref{32}), we are left to
show that $(1/2) \, [e_1{}^b(x),  \pi^a(x) \varepsilon_{ab} E(x)]_+ =
(1/2) \,\pi^a(x) \,\varepsilon_{ab}  [e_1{}^b(x), E(x)]_+$, but this
is true because of (\ref{60}).
\end{sloppypar}

\section{Outlook}

To carry through the Dirac quantization there are still two main problems
to be solved at this stage. The first one is the appearance of a term
proportional to $\delta (0)$  when one solves (\ref{61}) with hermitean $G_A$
by representing $e_1{}^b$ as a functional derivative operator.
The other one is the well--known problem of 'frozen time' (cf. Isham$^{12}$):
Due to (\ref{61}) $H_P$ vanishes on all physical states
so that a normal Schroedinger equation does not make sense.
These problems, among others, are treated in Ref.\,7.
--- What makes the quantization of the present model interesting is that
 there  appear similar conceptual problems as in $d = 4$, but that
the mathematical difficulties are much simpler to be overcome.
Therefore it is possible to explicitely check some of the
basic approaches to quantum gravity.

Another promising step seems to be the coupling of matter fields to the
action (\ref{1}), which, according to our results, is not restricted
by any  'hidden' symmetry.

\vspace{5ex}
{\Large\bf Acknowledgements}
\vspace{3ex}

I have profitted from numerous discussions with
W. Kummer, D. J. Schwarz, and especially P. Schaller, who also contributed the
idea to the simple proof following eq. (\ref{62}). I am also grateful to these
people, most of all W. Kummer, for many suggestions regarding the manuscript.

\begin{appendix}
\section{Attainability of the LC--Gauge}

%\renewcommand{\theequation}{\Alph{section}.\arabic{equation}}
%\doppelnummer

In this appendix we will prove the local attainability of the  gauge
\begin{eqnarray}
e_0{}^a &=& \delta (\hat{0}a)
\label{A1a}          \\                                    %A1a
\omega_0 &=& 0,
\label{A1b}                                                %A1b
\end{eqnarray}
which is the LC gauge as introduced in Kummer et. al.$^2$ when (\ref{A1a}) is
referred to the LC-metric (\ref{A2}). We will show further that (\ref{A1a}),
(\ref{A1b}) cannot be obtained for the topology of a torus.

As can be shown easily, the attainability of  (\ref{A1a}), (\ref{A1b})
is independent of the
choice of the reference metric $g_{ab}$. For convenience we will restrict
ourselves to (\ref{A2})  since the corresponding  local 'Lorentz'
transformations can be parametrized in the simple form $e_{\bar{a}}{}^b = {\rm
diag}(\exp(\alpha),\exp(-\alpha))$ with $\alpha =
\alpha(x) \in R$, where we have written simply $x$ for $x^\mu$
(in contrast to sec.\,3).
According to (\ref{54}), which is  certainly also valid when using (\ref{A2}),
 the complete gauge freedom can be expressed as
\begin{eqnarray}
\bar{\omega}_{\tilde{\mu}} (\tilde{x}) &=&
\frac{\partial x^\nu}{\partial \tilde{x}^\mu} (\tilde{x}) \:
(\omega_\nu - \alpha,_\nu) (x(\tilde{x}))
\label{A3}              \\                                            %A3
e_{\tilde{\mu}}{}^{\bar{a}} (\tilde{x}) &=&
\frac{\partial x^\nu}{\partial \tilde{x}^\mu} \,
e_b{}^{\bar{a}} \, e_\nu{}^b \,  (x(\tilde{x})),
\label{A4}                                                            %A4
\end{eqnarray}
with $e_a{}^{\bar{b}}$ being the inverse to $e_{\bar{a}}{}^b$.

Now, (\ref{A1a}) is equivalent to $e_{\hat{0}}{}^\mu =
\delta(0\mu)$. This in turn implies that   one has chosen the
flow of the vector field $e_{\hat{0}}$ as a coordinate $x^0$, and this  is
always possible.  Plugging (\ref{A1a}) into both sides of
(\ref{A4}), one gets the residual gauge freedom available to obtain
$\bar{\omega}_{\tilde{0}} = 0$ in a second step.  In this way it is
straightforward to show (using $e \neq 0$) that (\ref{A1b}) is attainable under
the assumption that  (\ref{A1a}) is already fulfilled, iff
\begin{eqnarray}
x^1 &=& x^1 (\tilde{x}^1)
\label{A5}              \\                                            %A5
\frac{\partial x^0}{\partial \tilde{x}^0} \,(\tilde{x}(x)) &=&
\exp (\alpha (x))
\label{A6}            \\                                              %A6
\alpha,_0(x) &=& \omega_0(x).
\label{A7}                                                            %A7
\end{eqnarray}
Because of (\ref{A5}) the lefthand side of (\ref{A6}) is just
$1/(\partial \tilde{x}^0/\partial x^0)(x)$ so that (\ref{A6},\ref{A7}) are
 solved by
\begin{eqnarray}
\alpha(x) &=& \int_{f(x^1)}^{x^0} dy \: \omega_0 (y,x^1)
\label{A8}                                                 \\           %A8
\tilde{x}^0 &=& \int_{g(x^1)}^{x^0} dy \, \exp (-\alpha(y,x^1))
\label{A9}                                                              %A9
\end{eqnarray}
with some arbitrary functions $f$ and $g$.

In a manifold with topology $S^1 \times S^1$ it is  possible to use just one
chart when requiring that all fields are periodic in $x^0$ and $x^1$ (e.g. with
the period normalized  to $2\pi$).  One concludes from (\ref{A7}) by
integrating this equation over a period in $x^0$ that even if
 (\ref{A1a}) is obtainable globally (\ref{A1b})  is not.

\end{appendix}

%\newpage                                                  %!!!
\vspace{5ex}                                             %!!!
{\Large\bf References}
\vspace{3ex}

\begin{enumerate}
  \item M. O. Katanaev and I. V. Volovich, {\em Phys. Lett.} {\bf B175},
                   413 (1986); {\em Ann. Phys.} (N.Y.) {\bf 197}, 1 (1990);
                  M. O. Katanaev, {\em J. Math. Phys.} {\bf 31}, 882 (1990).
  \item W. Kummer and D. J. Schwarz, {\em Phys. Rev.} {\bf D45}, 3628 (1992).
  \item M. O. Katanaev, {\em J. Math. Phys.} {\bf 32}, 2483 (1991);
                   M. O. Katanaev, {\em All universal coverings of
                   two--dimensional gravity with torsion}, preprint.
  \item   H. Grosse, W. Kummer, P. Pre$\check{\rm s}$najder and D. J.
                    Schwarz,  {\em  Novel Symmetry of Non--Einsteinian Gravity
                    in Two Dimensions}, preprint TUW--92--04, to appear in
                    {\em J. Math. Phys.} {\bf 33} (1992).
  \item M. Henneaux, {\em Phys. Rep.} {\bf 126}, 1 (1985).
  \item M. Henneaux, C. Teitelboim and  J. Zanelli, {\em Nucl. Phys.}
                   {\bf B332}, 169 (1990).
  \item  P. Schaller and T. Strobl, {\em Canonical Quantization of
                   Non--Einsteinian Gravity and the Problem of Time},
                   preprint TUW--92--13, hep-th/9211054.
  \item F. Haider, private communication.
  \item W. Kummer and D. J. Schwarz, {\em Renormalization of
                 $R^2$--Gravity with dynamical torsion in d=2},
                 preprint TUW--91--09, {\em Nucl. Phys. B}, to be published.
  \item I. A. Batalin and G. A. Vilkovisky, {\em Phys. Lett.}
                {\bf B55}, 224 (1975).
  \item K. Sundermeyer, {\em Constrained Dynamics},
                Lecture Notes in Physics {\bf 169},
                Springer Berlin Heidelberg New York 1982.
%  \item J. Govaerts, {\em Hamiltonian Quantisation and Constrained
 %                Dynamics}, Leuven University Press, 1991.
  \item C. J. Isham in {\em Recent Aspects of Quantum Fields},
                 ed. Gausterer et. al.
                 Lecture Notes in Physics {\bf 396}, p.123,
                 Springer Berlin Heidelberg 1991.
  \item P. A. M. Dirac, {\em Lectures on Quantum Mechanics},
                 Yeshiva University, New York 1964.
  \item M. J. Gotay, {\em J. Phys.} {\bf A16}, L141 (1983).
\end{enumerate}

\end{document}